\documentclass[12pt,a4paper]{article}
\usepackage{latexsym}
\DeclareSymbolFont{AMSb}{U}{msb}{m}{n}
\DeclareSymbolFontAlphabet{\Bbb}{AMSb}
\newcommand{\N}{\Bbb{N}}
\newcommand{\Z}{\Bbb{Z}}
\newcommand{\R}{\Bbb{R}}

\newtheorem{proposition}{Proposition}
\newcommand{\beq}{\begin{equation}}
\newcommand{\eeq}{\end{equation}}

\newcommand{\proof}{\textbf{Proof.\ }}
\newcommand{\qed}{\begin{flushright} $\Box$ \end{flushright}}

\begin{document}
\thispagestyle{empty}
\vspace*{-80pt} 
{\hspace*{\fill} Preprint-KUL-TF-98/53} 
\vspace{80pt} 
\begin{center} 
{\LARGE Goldstone Boson Normal Coordinates \\[10pt] in Interacting Bose Gases  }
 \\[25pt]  
 
{\large  
    T.~Michoel\footnote{Aspirant van het Fonds voor Wetenschappelijk
    Onderzoek - Vlaanderen}\footnotetext{Email: {\tt tom.michoel@fys.kuleuven.ac.be}},
	A.~Verbeure\footnote{Email: {\tt andre.verbeure@fys.kuleuven.ac.be}}
    } \\[40pt]   
{Instituut voor Theoretische Fysica} \\  
{Katholieke Universiteit Leuven} \\  
{Celestijnenlaan 200D} \\  
{B-3001 Leuven, Belgium}\\[75pt]
\end{center} 
\begin{abstract}\noindent
For the phenomenon of Bose-Einstein condensation we construct the canonical pair of
field operators of the Goldstone Bosons explicitly as fluctuation operators in the ground
state. We
consider the imperfect Bose gas as well as the weakly interacting Bose gas. We prove
that a canonical pair of fluctuation operators is always related to the order
parameter and the generator of the broken symmetry fluctuations. We find that although
the first one has an anomalous behaviour, the second one is squeezed by the same inverse
rate. Furthermore, we prove that this canonical pair separates from the other variables of
the system and that it behaves dynamically as oscillator variables. Finally the long
wavelength behaviour of the spectrum determines the lifetime of this pair.
\end{abstract}
\newpage

\section{Introduction}
The phenomenon of spontaneous symmetry breaking (SSB) is a representative tool
for the explanation of many phenomena in modern physics of field theory and
statistical mechanics of many-body theory. The analysis of SSB goes back to the 
Goldstone Theorem \cite{1}, which has been the subject of much analysis. It is
proved that for short range interactions in many-body systems SSB implies the
absence of an energy gap in the excitation spectrum \cite{2,3}. 

For long range interactions the SSB has also been studied extensively. 
In the physics literature the phenomenon is known as the occurence of 
oscillations with frequency spectrum taking a finite value $\omega \not= 0$ at 
$k=0$ \cite{4,5,6}. Different approximation methods, typical here is the 
random phase approximation, yield the exact computation of these frequencies. 
For some mean field models, the BCS-model \cite{7}, the Overhauser model
\cite{8}, the anharmonic crystal model \cite{9}, and for the Jellium model
\cite{10}, we were able to give the mathematical status of these frequencies as
elements of the spectrum of typical fluctuation operators (see \cite{11,12}).

The typical operators entering in the discusssion are the generator of the
broken symmetry and the order parameter. In physical terms expressed, it is the
charge or density operator and the current operator. Their fluctuation operators
form a quantum canonical pair, which decouples from the other degrees of freedom
of the system. As fluctuation operators are collective operators, they describe
the collective mode accompanying the SSB phenomenon. Hence for long range
interacting systems, we realised mathematically rigorously in these models, the
so-called Anderson theorem \cite{13,14} of `restauration of symmetry',
 stating that there exists a spectrum
of collective modes $\omega(k\to 0)\not= 0$ and that the mode in the limit 
$k\to 0$ is the operator which connects the set of degenerate temperature
states, i.e. `rotates' one ergodic state into an other. We conjecture that our
results of \cite{7,8,10} can be proved for general long range two-body
interacting systems as a universal theorem.
Anderson did formulate his theorem in the context of the Goldstone theorem for
short range interacting systems, i.e. in the case $\omega(k\to 0)=0$ of absence
of an energy gap in the ground state. 

Of course one knows that there is no
one-to-one relation between short range interactions and the absence of an
energy gap for symmetry breaking systems (see e.g. \cite{9}). The imperfect
Bose gas is an example of a long range interacting system showing SSB, but
without energy gap. In this paper we realise the above described program of
construction of the collective modes operators of condensate density and
condensate current, as normal modes dynamically independent from the other
degrees of freedom of the system. We consider the whole temperature range, the
ground state included.

In particular the ground state situation is interesting, because it yields a
non-trivial quantum mechanical canonical pair of conjugate operators, giving an
explicit representation of the field variables of the socalled Goldstone boson. One
can consider this result as a formal step forward beyond the known analysis of the
Goldstone phenomenon.

Moreover in Section \ref{sectionwibg}, we extend this result tot the weakly
interacting Bose gas of superfluidity. It is interesting to remark that here the
situation is intrinsically different in the sense that only the condenste density mode
is spontaneously broken. One checks explicitly that the density fluctuation operator
and the order parameter fluctuation operator do not form a non-trivial pair, but the
condensate density and the order parameter fluctuation operators do.

Hence in both models the fluctuation operators of the generator of the broken symmetry
and of the order parameter form a non-trivial canonical pair. The latter one
shows off-diagonal long range order, therefore the density-density correlation can not
share this property. This can be interpreted as that a spontaneously broken symmetry
behaves like an approximate symmetry. The explicit construction of the canonical pair
amounts to the realisation of `restauration of symmetry', an idea put forward by Anderson
\cite{13,14}.

Furthermore, for both models, we prove that the canonical pair of Goldstone fluctuation
modes separates dynamically from the other variables of the system and behaves like
harmonic oscillator modes with a frequency proportional to the condensate density, i.e.
this phenomenon disappears if no condensation is present. It turns out that this pair of
variables has a lifetime in the long wavelength limit which is determined by the long
wavelength behaviour of the spectrum of the system.

\section{Fluctuation operators}\label{flucop}
We want to study different models of a Bose gas in which there is breaking of
the gauge symmetry. In general, a system of identical bosons of mass $m$ in a
cubic box $\Lambda \subset \R^\nu$ of volume $V=L^\nu$, $\nu\geq 3$
with periodic boundary
conditions for the wave functions, is described by the full two-body
interaction Hamiltonian
\beq\label{genH}
H_L = \sum_k \epsilon_k a^*_{L,k}a_{L,k} + \frac{1}{2V}\sum_{q,k,k'}
v(q)a^*_{L,k+q}a^*_{L,k'-q}a_{L,k'}a_{L,k} - \mu_L N_L,
\eeq 
where the sum runs over the set $\Lambda^*=\frac{2\pi}{L}\Z^\nu$ and 
$\epsilon_k=
\frac{|k|^2}{2m}$; $a^\sharp_{L,k}$ are the boson creation/annihilation
operators in the one-particle state $\psi_{L,k}(x)=V^{-1/2}e^{ik.x}$,
$x\in\Lambda$, $k\in\Lambda^*$, i.e.
\beq
a_{L,k}=\int_\Lambda  a(x)\frac{e^{-ik.x}}{V^{1/2}} dx,
\eeq
\[
[a(x),a^*(y)]=\delta(x-y)
\]
and
$v(q)=\int_{\R^\nu} e^{-iq.x}\phi(x) dx$, $\phi$ is the periodically extended
two-body interaction potential.

The generator of the gauge symmetry is the total number operator $N_L$, with
generator density $a^*(x)a(x)$:
\[
N_L = \int_{\Lambda}a^*(x)a(x) dx.
\]
The common choice of order parameter is $V^{-1/2}a^\sharp_{L,0}$, or taking a
self-adjoint combination 
\[
A_L=\frac{i}{\sqrt{2V}}(a^*_{L,0}-a_{L,0})=\frac{i}{\sqrt{2}V}\int_{\Lambda}
 (a^*(x)-a(x)) dx,
\]
so the order parameter density is given by $\frac{i}{\sqrt{2}}(a^*(x)-a(x))$.
One has of course.
\[
[N_L,A_L]=\frac{i}{\sqrt{2V}}(a^*_{L,0}+a_{L,0})
\stackrel{{V\to\infty}}{\longrightarrow}i\sqrt{2\rho_0}\cos \alpha,
\]
where $\rho_0$ is the density of the condensate and $\alpha$ the phase,i.e.
\[
V^{-1/2}a^*_{L,0} \to \sqrt{\rho_0}e^{i\alpha}.
\]

We are here  interested in the behaviour of the $q$-mode fluctuation
($q\not= 0$) \cite{14b} of this generator and order parameter, i.e.
\begin{eqnarray}
F_{L,q}(N) &=& \frac{1}{V^{1/2}}\int_{\Lambda} a^*(x)a(x)e^{iq.x} dx\\
F_{L,q}(A) &=& \frac{i}{\sqrt{2}V^{1/2}}\int_{\Lambda}(a^*(x)-a(x))e^{iq.x}dx,
\end{eqnarray}
which satisfy the same commutation relation as $N_L$ and $A_L$:
\beq
[F_{L,q}(N),F_{L,-q}(A)]=\frac{i}{\sqrt{2V}}(a^*_{L,0}+a_{L,0})
\stackrel{{V\to\infty}}{\longrightarrow}i\sqrt{2\rho_0}\cos\alpha.
\eeq
In fact we take a sequence $0\not= q_L \in \Lambda^*$ converging to $q$ so that
there is no need to substract expectation values, since for $q\in\Lambda^*$,
$\int_{\Lambda}e^{iq.x}dx=V\delta_{q,0}$, and so that we can also write
\begin{eqnarray}
F_{L,q}(N) &=& \frac{1}{V^{1/2}}\sum_k a^*_{L,k+q}a_{L,k}\\
F_{L,q}(A) &=& \frac{i}{\sqrt{2}}(a^*_{L,q}-a_{L,-q}).
\end{eqnarray}

Our first goal will be to define these operators in the thermodynamic
limit $L\to\infty$. This will be done via a central limit theorem, as defined
in \cite{11,12}. Afterwards we will be interested in the long wavelength - low
frequency limit $q\to 0$ in which collective behaviour is to be expected. 
In this limit $q\to 0$ we give a connection with the abstractly studied
fluctuation operators \cite{14c} of the type
\beq\label{deltafluc}
F_\delta(O)=\lim_{L\to\infty}F_{L,\delta}(O)=\frac{1}{V^{\frac{1}{2}+\delta}}
\int_{\Lambda}(O(x)-<O(x)>)dx,
\eeq
where $O$ is some operator density and $\delta$ a critical exponent describing the
degree of abnormality of the fluctuations of $O$, defined by existence of the
variance. If $\delta >0$ there is ODLRO, if $\delta < 0$, the fluctuation is squeezed. 

In an interacting Bose gas this $q\to 0$ behaviour will be mainly determined by
the spectrum $E_q$ of the Hamiltonian.
The density fluctuation $F_{L,q}(N)$ has another very important property, its
commutator with the two-body interaction part of the Hamiltonian vanishes:
\beq\label{[U,F]}
[U_L,F_{L,q}(N)]=0,
\eeq
where
\[
U_L=\frac{1}{2V}\sum_{q,k,k'}v(q)a^*_{L,k+q}a^*_{L,k'-q}a_{L,k'}a_{L,k}.
\]
This is easily seen as follows. Commute in $U_L$, $a^*_{L,k'-q}$
with $a_{L,k}$. This gives
\[
U_L=\frac{1}{2V}\sum_{q,k,k'}v(q)a^*_{L,k+q}a_{L,k}a^*_{L,k'-q}a_{L,k'}
- \frac{1}{2V}\sum_{q,k}v(q)a^*_{L,k+q}a_{L,k+q}.
\]
In the first term, separate the term $q=0$ from the rest, use translation
invariance in the second term, and observe that $U_L$ can then be written as
\beq\label{U}
U_L=\frac{1}{2}\sum_{q\not=0}v(q)F_{L,q}(N)F_{L,-q}(N) + 
\frac{v(0)}{2V}N_L^2-\frac{1}{2}\phi(0)N_L.
\eeq
From this expression, (\ref{[U,F]}) is obvious.

In physics, one encounters essentially two types of Bose condensed systems,
namely those with a quadratic excitation spectrum ($E_q\propto |q|^2$, $|q|$
small) and those with a superfluid, linear spectrum ($E_q\propto |q|$, $|q|$
small). We will treat in detail an example of each of these cases.

\section{The Imperfect Bose Gas}
\subsection{The model and equilibrium states}
To make things more concrete, we consider as a first example the imperfect or
mean field Bose gas \cite{15,16}, specified by the local
Hamiltonian $H_L$ with periodic boundary conditions \cite{17}:
\begin{equation}\label{H}
H_L = T_L - \mu_L N_L + \frac{\lambda}{2 V} N_L^2
\end{equation}
where $\lambda \in \R^+$, $\Lambda$ the centered cubic box of side length $L$ in
$\R^\nu$, $\nu\geq 3$, $\Lambda^*=\frac{2\pi}{L}\Z^\nu$.
Remark that, apart from a shift in the chemical potential, this Hamiltonian 
can be obtained from (\ref{genH}) by taking $v(q)=0$ if $q\not= 0$ in (\ref{U}).

Talking about the thermodynamic limit, we mean $L\to\infty$ under the constraint
that for all $L$
\beq
\frac{\omega_L(N_L)}{V}=\rho,
\eeq
where $\rho$ is any positive number standing for the average density of
particles, $\omega_L$ is the canonical Gibbs state for (\ref{H}) at some inverse
temperature $\beta$. It is proved \cite{17} that $\omega_\beta(.)=\lim_L 
\omega_L(.)$ exists as a space homogeneous state on the algebra of polynomials
in  the creation and annihilation operators. It is proved that there exists
condensation in the zero ($k=0$) mode state if $\rho$ is large enough and $T$ is
small enough.

The phase transition is accompanied by a spontaneous breaking of the gauge
symmetry of (\ref{H}) in the sense that
\beq\label{ssb}
\lim_{L\to\infty}\omega_L\left( \frac{a^*_{L,0}}{V^{1/2}}a(f) \right)
= \rho_0 \hat f(0) \; , \; \rho_0>0,
\eeq 
where
$
a(f)=\int  \overline{f(x)}a(x) dx
$
for $f\in C^\infty_0(\R^\nu)$,
$\hat f$ is the Fourier transform of $f$ and $\rho_0$ is the condensate density.
It is proved in \cite{18} that (\ref{ssb}) implies, amongst other things, the
breaking of the gauge symmetry.

The limit chemical potential is given by
\beq
\mu = \lambda \rho,
\eeq
and the dynamics coincides with the dynamics of the free Bose gas.
The limit Gibbs state has the following form: for all local observables $A$,
\beq
\omega_\beta(A) = \frac{1}{2\pi}\int_0^{2\pi} \omega_\beta^\alpha(A) d\alpha,
\eeq
with
\beq\label{state}
\omega_\beta^\alpha\left( e^{i(a(f) + a^*(f))}  \right)
= \exp\left[ -\frac{1}{2}(f,Kf) + 2i\rho_0^{1/2}|\hat f(0)| \cos \alpha \right]
\eeq
and
\beq
\left( \widehat{Kf} \right) (k) = \frac{1}{2} \coth(\frac{\beta \epsilon_k}{2})
\hat f (k).
\eeq

The states $\omega_\beta^\alpha (\alpha\in [0,2\pi] )$ are the extremal
equilibrium state components of $\omega_\beta$ with the property that
\beq
\lim_{L\to\infty}\omega_\beta^\alpha \left(\frac{a^*_{L,0}}{V^{1/2}} \right) =
\sqrt{\rho_0}e^{i\alpha},
\eeq
and as operators in the GNS-representation of $\omega_\beta^\alpha$, one has
also
\beq\label{a0}
\lim_{L\to\infty}\frac{a^*_{L,0}}{V^{1/2}} =
\sqrt{\rho_0}e^{i\alpha}.
\eeq
Remark also that the states $\omega_\beta^\alpha (\alpha\in [0,2\pi] )$ are
quasi-free states, making the computation of expectation values straightforward.

\subsection{Collective Goldstone modes}
We now turn our attention to the density and order parameter fluctuations. 
We consider our system to be in one of the extremal equilibrium states
$\omega_\beta^\alpha$, for some $\alpha\in [0,2\pi]$, and without loss of
generality, we take $\alpha=0$, and denote this state again by $\omega_\beta$.

For notational convenience, if $\rho_0\not= 0$, denote
 \[
\rho_{L,q}=\frac{1}{\sqrt{2\rho_0}}F_{L,q}(N), \; \;\;\;A_{L,q}=F_{L,q}(A).
\]
Then we have
\[
[\rho_{L,q},A_{L,-q}]=\frac{i}{2\sqrt{\rho_0 V}}(a^*_{L,0} + a_{L,0}),
\]
and by (\ref{a0}):
\beq
\lim_{L\to\infty}[\rho_{L,q},A_{L,-q}]=i.
\eeq
More generally
\beq
\lim_{L\to\infty}[\rho_{L,q},A_{L,-q'}]=i\delta_{q,q'}.
\eeq

Let us first calculate the variances of $\rho_{L,q}$ and $A_{L,q}$.
\begin{proposition}
We have for $q,q'\not=0$
\begin{enumerate}
\item
\[
\lim_{L\to\infty}\omega_\beta(\rho_{L,q}\rho_{L,-q'})
\]
\[
= \delta_{q,q'}\left(
\frac{1}{2}\coth\frac{\beta\epsilon_q}{2}   +\frac{1}{2\rho_0} \int_{\R^\nu}
\frac{dk}{(2\pi)^\nu}\frac{1}{e^{\beta\epsilon_{k+q}}-1} \frac{1}{1-e^{-\beta
\epsilon_k}} \right),
\]

\item
\[
\lim_{L\to\infty}\omega_\beta(A_{L,q}A_{L,-q'}) = \delta_{q,q'}\frac{1}{2}
\coth\frac{\beta\epsilon_q}{2}.
\]
\end{enumerate}
\end{proposition}
\proof
The proof is a straightforward calculation using the quasi-freeness of the state
$\omega_\beta$, e.g.
\begin{eqnarray*}
&&\omega_\beta(\rho_{L,q}\rho_{L,-q})\\
 &=& \frac{1}{2\rho_0 V}\sum_{k,k'}
\omega_\beta(a^*_{L,k+q}a_{L,k}a^*_{L,k'-q}a_{L,k'}) \\
&=& \frac{1}{2\rho_0 V}\sum_{k}\omega_\beta(a^*_{L,k+q}a_{L,k+q})
\omega_\beta(a_{L,k}a^*_{L,k})\\
&=& \frac{1}{2\rho_0 V}\left( \omega_\beta(a^*_{L,q}a_{L,q})
\omega_\beta(a_{L,0}a^*_{L,0}) + \omega_\beta(a^*_{L,0}a_{L,0})
\omega_\beta(a_{L,-q}a^*_{L,-q}) \right) \\
&& + \frac{1}{2\rho_0 V}\sum_{-q\not=k\not=0}\omega_\beta(a^*_{L,k+q}a_{L,k+q})
\omega_\beta(a_{L,k}a^*_{L,k}).
\end{eqnarray*}
In the limit, this becomes
\begin{eqnarray*}
\lim_{L\to\infty}\omega_\beta(\rho_{L,q}\rho_{L,-q}) &=& \frac{1}{2}\coth\frac{\beta
\epsilon_q}{2}+ \frac{1}{2\rho_0} \int_{\R^\nu}
\frac{dk}{(2\pi)^\nu}\frac{1}{e^{\beta\epsilon_{k+q}}-1} \frac{1}{1-e^{-\beta
\epsilon_k}}.
\end{eqnarray*}
The other case is even easier.
\qed
From this already a few conclusions can be drawn. First of all, consider
 the integral in the most relevant case $\nu=3$:
\[
\frac{1}{2\rho_0} \int_{\R^3}
\frac{dk}{(2\pi)^3}\frac{1}{e^{\beta\epsilon_{k+q}}-1} \frac{1}{1-e^{-\beta
\epsilon_k}}.
\]
Letting $q\to 0$, this integral clearly diverges due to the contribution of the
neighbourhood of $k=0$. Near $k=0$ we can write it (up to constants) like
\[
\int\frac{dk}{(2\pi)^3}\frac{1}{(k+q)^2 k^2}.
\]
Taking $q$ e.g. along the $z$-axis and changing the variable $k$ to 
$k'=\frac{k}{|q|}$, it can be seen that this integral diverges like $|q|^{-1}$.
Since $\coth\frac{\beta\epsilon_q}{2}$ diverges as $|q|^{-2}$ for $q\to 0$, 
we see that for small $q$ the
variance of $\rho_{L,q}$ is completely dominated by the $\coth$-term.

This divergence implies that we should renormalise both $\rho_{L,q}$ and $A_{L,q}$ in
order to get a nontrivial limit $q\to 0$ of their variances, i.e.
\begin{eqnarray*}
\rho_{L,q} &\to& \tilde \rho_{L,q}=|q|\rho_{L,q},\\
A_{L,q} &\to& \tilde A_{L,q}=|q|A_{L,q}.
\end{eqnarray*}
But this implies that the commutator
\[
\lim_{L\to\infty}[\tilde \rho_{L,q},\tilde A_{L,-q}]=i|q|^2,
\]
vanishes in the limit $q\to 0$. 

On the other hand, if one considers the ground state situation (limit $\beta\to\infty$),
the $q\to 0$ analysis yields that the variances
\[
\lim_{q\to 0}\omega_\infty(\rho_{L,q}\rho_{L,-q})
\]
and
\[
\lim_{q\to 0}\omega_\infty(A_{L,q}A_{L,-q})
\]
are both finite, and the commutation relation between $\rho_{L,q}$ and $A_{L,-q}$ is
non-trivial and canonical.

This is not surprising, as one expects true quantum effects on the level of fluctuations
only in the ground state. Critical quantum effects are hidden behind the temperature
($T>0$) fluctuations.

Moreover, it is tempting to identify this renormalisation in $q$ with the exponent
$\delta$ of (\ref{deltafluc}) via the relation $|q|\propto L^{-1}$. This relation of 
course being given by
the fact that the first non-zero $q$-level in finite volume is $|q|=2\pi
L^{-1}$. In that case we obtain for the density fluctuations $F_{L,q}(N)$ that
$\delta=1/3$ in the condensed phase. In the normal phase, the $\coth$-term would be
absent and the integral would be convergent also for $q=0$ because $e^{\beta
\epsilon_k}$ would be replaced by $e^{\beta(\epsilon_k - \alpha)}$, with $\alpha<0$,
hence $\delta= 0$.
At the critical point, the $\coth$-term would still be absent but the integral would now
be divergent like $|q|^{-1}$ as shown before. This would then give $\delta=1/6$. These
three values for $\delta$ are exactly the ones calculated in \cite{18b}.

Since we will be interested in quantum effects on the level of macroscopic fluctuations 
we will restrict ourself from now on to the ground state (from now on denoted $\omega$).
We redefine $\rho_{L,q}$ and $A_{L,q}$ as self-adjoint operators, and we make
the (arbitrary) choice of taking the \emph{cos-fluctuation}:
\begin{eqnarray}
\rho_{L,q} &=& \frac{1}{\sqrt{\rho_0 V}}\int_\Lambda  a^*(x) a(x) \cos(q.x) dx\\
A_{L,q} &=& \frac{i}{\sqrt{V}}\int_\Lambda  (a^*(x) - a(x)) \cos(q.x) dx,
\end{eqnarray}
or in momentum space
\begin{eqnarray}
\rho_{L,q} &=& \frac{1}{2\sqrt{\rho_0 V}}\sum_k (a^*_{L,k+q}a_{L,k}+a^*_{L,k-q}a_{L,k})
\\
A_{L,q} &=& \frac{i}{2}[a^*_{L,q}+a^*_{L,-q}-(a_{L,q}+a_{L,-q})],
\end{eqnarray}
where the normalization is chosen such that
\beq
\lim_{L\to\infty}[\rho_{L,q}, A_{L,q'}]=i\delta_{q,q'}.
\eeq

It is easy to check that with these definitions  
\begin{eqnarray}
\lim_{L\to\infty}\omega(\rho_{L,q}^2) &=& \frac{1}{2} \\
\lim_{L\to\infty}\omega(A_{L,q}^2) &=& \frac{1}{2}.
\end{eqnarray}

The rest of this section is devoted to the more mathematical aspects of  the realisation
of the different fluctuation operators as central limits of operators. The less
mathematics minded reader can skip this part at a first reading and proceed immediately to
section \ref{dynamics-1}.

Let $\mathcal{F}$ be the family of complex continuous functions $f(k,k')$ 
of two variables $k,k' \in \R^\nu$, satisfying
\beq
f(\pm k, \pm k')=f(k,k')
\eeq
and
\beq\label{complex-cond}
\overline{f(k,k')}=f(k',k).
\eeq
 With later applications in mind, define for $f,g \in\mathcal{F}$,
$\rho_{L,q}(f)$ and $A_{L,q}(g)$ by
\begin{eqnarray}
\rho_{L,q}(f) &=& \frac{1}{2\sqrt{\rho_0 V}}\sum_k \left[f(k+q,k)a^*_{L,k+q}a_{L,k}
\right.\\
&&\left.+ f(k-q,k)a^*_{L,k-q}a_{L,k}\right]\nonumber\\
A_{L,q}(g) &=& \frac{i}{2}\left[g(q,0)(a^*_{L,q}+a^*_{L,-q})-g(0,q)(a_{L,q}+a_{L,-q})
\right].
\end{eqnarray}
Condition (\ref{complex-cond}) ensures the self-adjointness of these operators.
Then
define operators $F_{L,q}(f,g)$ by
\beq
F_{L,q}(f,g) =\rho_{L,q}(f) +  A_{L,q}(g).
\eeq

\begin{proposition} For $f,g\in\mathcal{F}$,
\beq
\lim_{L\to\infty}\omega\left( F_{L,q}(f,g)^2 \right)
= \frac{1}{2}\left| f(q,0) + i g(q,0) \right|^2.
\eeq
\end{proposition}
\proof
This is a simple calculation using the quasi-freeness of the state $\omega$.
\qed

In the GNS-representation $(\mathcal{H}_\omega, \pi_\omega, \Omega_\omega)$ of the state 
$\omega$ a scalar product is defined by
\[
<\pi_\omega(A)\Omega_\omega,\pi_\omega(B)\Omega_\omega>_\omega=\omega(A^*B),
\] 
and the associated norm is denoted by $\|.\|_\omega$.

Denoting $\pi_\omega(F_{L,q}(f,g))$ again by $F_{L,q}(f,g)$, we then have:
\begin{proposition}[A BCH-formula]\label{BCH-prop}
Let $f_i,g_i \in \mathcal{F}, i=1,2$, then
\[
\lim_{L\to\infty}\left\| e^{iF_{L,q}(f_1,g_1)}e^{iF_{L,q}(f_2,g_2)} -
e^{i(F_{L,q}(f_1,g_1) + F_{L,q}(f_2,g_2))}
e^{-\frac{1}{2}[F_{L,q}(f_1,g_1),F_{L,q}(f_2,g_2)]} \right\|_\omega
\]
\beq\label{BCH}
 = 0.
\eeq
\end{proposition}
\proof 
See Appendix \ref{BCH-append}.
\qed
This result should be compared with the Baker-Campbell-Haussdorff formula which states that
for two operators $A,B$ whose commutator is a complex number:
\[
e^A e^B = e^{(A+B)}e^{-\frac{1}{2}[A,B]}.
\]
This proposition tells us that in a weak sense this BCH-formula remains true for our 
fluctuation
operators, whose commutator becomes a complex number in the thermodynamic limit. Since we
are studying fluctuations of \emph{unbounded} operators, the BCH-formula is only true in
the GNS-representation of $\omega$. For fluctuations of \emph{bounded} operators, the 
BCH-formula
holds in a much stronger sense, independent of the state (see \cite{14b}).

On the complex vectorspace $\mathcal{V}$ of complex linear combinations of elements from 
$(\mathcal{F},\mathcal{F})$,
define a sesquilinear form $<\cdot|\cdot>_q$ by
\begin{eqnarray}
<f_1,g_1|f_2,g_2>_q &=& \lim_{L\to\infty}\omega\left(F_{L,q}(f_1,g_1)^*
F_{L,q}(f_2,g_2)\right)\nonumber\\
&=& \frac{1}{2}\overline{\left(f_1(q,0)+ig_1(q,0)\right)}
\left(f_2(q,0)+ig_2(q,0)\right),
\end{eqnarray}
and extension to the whole of $\mathcal{V}$ by linearity.

This form is positive and satisfies the Cauchy-Schwarz inequality by the
positivity of $\omega$ and the Cauchy-Schwarz inequality for the state
$\omega$.

Separating the real and the imaginary part of the restriction of $<\cdot|\cdot>_q$ to
the real subspace  $(\mathcal{F},\mathcal{F})$ of $\mathcal{V}$, i.e.
\beq
<f_1,g_1|f_2,g_2>_q = s_q\left(f_1,g_1|f_2,g_2\right) + \frac{i}{2}
\sigma_q\left(f_1,g_1|f_2,g_2\right),
\eeq
defines a real bilinear positive symmetric form $s_q$ and a symplectic form
$\sigma_q$. (A form $\sigma$ is called symplectic if $\sigma(x,y)=-\sigma
(y,x)$.)

The symplectic form $\sigma_q$ satisfies
\beq
\lim_{L\to\infty}[F_{L,q}(f_1,g_1),F_{L,q}(f_2,g_2)]=i
\sigma_q\left(f_1,g_1|f_2,g_2\right),
\eeq
where the limit is taken in the GNS-representation of $\omega$.

The following proposition is the crucial Central Limit Theorem for the operators
$F_{L,q}(f,g)$.
\begin{proposition}[Central Limit Theorem]\label{cltprop}
For $f,g\in\mathcal{F}$, $t\in\R$,
\beq\label{clt}
\lim_{L\to\infty}\omega\left( e^{itF_{L,q}(f,g)} \right)
=e^{-\frac{t^2}{2}s_q\left(f,g|f,g\right)}.
\eeq
\end{proposition}
\proof Although a similar theorem could also be proven for temperature states, we
will only do it for the ground state, since that is really all we need. In that
case, using the quasi-freeness of the state and the fact that all particles are
condensed into the zero-energy state simplifies the proof. The details can be found
in Appendix \ref{CLT-proof}.
\qed

The $C^*$-algebra of the canonical commutation relations over $(H,\sigma)$, with
$H$ a real linear space and $\sigma$ a symplectic form, written as
$CCR(H,\sigma)$, is by definition a $C^*$-algebra generated by elements $\{W(f)
:f\in H\}$ such that
\begin{enumerate}
\item $W(-f)=W(f)^*$

\item $W(f)W(g)=e^{\frac{i}{2}\sigma(f,g)}W(f+g)$.

\end{enumerate}
Condition (ii) tells us that $W(f)W(0)=W(0)W(f)=W(f)$. Hence $W(0)$ is the unit of
the algebra and it follows that $W(f)$ is a unitary for every $f$. For an
elaborate discussion of the $CCR$, we refer to \cite{18c}.

\begin{proposition}[Reconstruction Theorem]\label{recthm}
The linear functional
\beq
\tilde \omega^q \left( W_q(f,g) \right) = e^{-\frac{1}{2}s_q(f,g|f,g)}
\eeq
defined on the algebra $CCR((\mathcal{F},\mathcal{F}),\sigma_q)$, is a quasi free state.

More explicitly, we have for all $(f_i,g_i) \in (\mathcal{F},\mathcal{F}), 
i=1,\ldots,n$,
\beq\label{clt-expl}
\lim_{L\to\infty}\omega\left(  e^{iF_{L,q}(f_1,g_1)}\cdots
 e^{iF_{L,q}(f_n,g_n)}\right) = \tilde \omega^q \left( W_q(f_1,g_1)
\cdots W_q(f_n,g_n) \right).
\eeq

The state $\tilde \omega^q$ is regular and hence for every $(f,g)$ there
exists a self-adjoint Bosonic field $\Phi_q(f,g)$ in the GNS representation
$(\mathcal{H}_{\tilde \omega^q},\pi_{\tilde \omega^q},
\Omega_{\tilde \omega^q})$ such that
\beq
\pi_{\tilde \omega^q}\left( W_q(f,g) \right) = e^{i\Phi_q(f,g)}.
\eeq

This implies that in the sense of the central limit (\ref{clt-expl}), the local
fluctuations converge to the Bosonic fields associated with 
$CCR((\mathcal{F},\mathcal{F}),\sigma_q)$:
\beq
\mathrm{CLT}-\lim_{L\to\infty}F_{L,q}(f,g)=\Phi_q(f,g).
\eeq
\end{proposition}
\newpage
\proof
See Appendix \ref{recthm-proof}.
\qed

The following definitions now clearly make sense:
\begin{eqnarray}
\rho_q &=& \Phi_q(1,0)=\mathrm{CLT}-\lim_{L\to\infty}\rho_{L,q},\\
A_q &=& \Phi_q(0,1)=\mathrm{CLT}-\lim_{L\to\infty}A_{L,q}.
\end{eqnarray}

In the same spirit as this central limit, we now define a limit 
$q\to 0$ of the operators $\Phi_q(f,g)$.

Define a sesquilinear form $<\cdot|\cdot>$ on $\mathcal{V}$ by
\beq
<f_1,g_1|f_2,g_2> = \lim_{q\to 0}<f_1,g_1|f_2,g_2>_q,
\eeq
and a real linear form $s$ and a symplectic form $\sigma$ in the obvious way:
\begin{eqnarray}
s\left(f_1,g_1|f_2,g_2\right)&=&\lim_{q\to 0}s_q\left(f_1,g_1|f_2,g_2\right)\\
\sigma\left(f_1,g_1|f_2,g_2\right)&=&\lim_{q\to 0}\sigma_q
\left(f_1,g_1|f_2,g_2\right).
\end{eqnarray}

We then get the limit ($q\to 0$) result:
\begin{proposition}[Reconstruction Theorem 2]\label{recthm2}
The linear functional
\beq
\tilde \omega\left( W(f,g) \right)=\lim_{q\to 0}
\tilde \omega^q \left( W_q(f,g) \right) = e^{-\frac{1}{2}s(f,g|f,g)}
\eeq
defined on the algebra $CCR((\mathcal{F},\mathcal{F}),\sigma)$, is a quasi free state.

More explicitly, we have for all $(f_i,g_i) \in (\mathcal{F},\mathcal{F}), 
i=1,\ldots,n$,
\beq\label{clt-expl2}
\lim_{q\to 0}\lim_{L\to\infty}\omega\left(  e^{iF_{L,q}(f_1,g_1)}\cdots
 e^{iF_{L,q}(f_n,g_n)}\right) = \tilde \omega \left( W(f_1,g_1)
\cdots W(f_n,g_n) \right).
\eeq

The state $\tilde \omega$ is regular and hence for every $(f,g)$ there
exists a self-adjoint Bosonic field $\Phi(f,g)$ in the GNS representation
$(\mathcal{H}_{\tilde \omega},\pi_{\tilde \omega},
\Omega_{\tilde \omega})$ such that
\beq
\pi_{\tilde \omega}\left( W(f,g) \right) = e^{i\Phi(f,g)}.
\eeq

This implies that in this sense of the limit $q\to 0$ (\ref{clt-expl2}), 
the fluctuations $\Phi_q$ converge to the Bosonic fields associated with 
$CCR((\mathcal{F},\mathcal{F}),\sigma)$:
\beq
\Phi(f,g)=\lim_{q\to 0}\Phi_q(f,g)=\mathrm{CLT}-\lim_{q\to 0}\lim_{L\to\infty}
F_{L,q}(f,g).
\eeq
\end{proposition}
\proof
This is just a matter of taking the limit $q\to 0$ in the different steps of the proof of
the previous Proposition.
\qed

Specifying again to our original operators:
\begin{eqnarray}
\tilde \rho&=&\lim_{q\to 0}\rho_q = \Phi(1,0)=\mathrm{CLT}-\lim_{q\to 0}
\lim_{L\to\infty}\rho_{L,q},\\
\tilde A &=&\lim_{q\to 0}A_q = \Phi(0,1)=\mathrm{CLT}-\lim_{q\to 0}
\lim_{L\to\infty}A_{L,q}.
\end{eqnarray}

The algebra of macroscopic fluctuations $CCR((\mathcal{F},\mathcal{F}),\sigma)$
 is a coarse grained one, i.e. different microscopic observables can have the
same macroscopic fluctuation operators. To describe this mathematically,
introduce an equivalence relation $\sim$ on $\mathcal{V}$ by
\beq
(f_1,g_1) \sim (f_2,g_2) \;\; \iff \;\; <f_1-f_2,g_1-g_2|f_1-f_2,g_1-g_2>=0.
\eeq
Another way of stating the equivalence relation is of course
\beq
(f_1,g_1) \sim (f_2,g_2) \;\; \iff \;\; 
\lim_{q\to 0}\lim_{L\to\infty}\omega\left( F_{L,q}(f_1-f_2,g_1-g_2)^2
\right)=0.
\eeq
We then have the following result:
\begin{proposition}\label{equiv-prop}
For $f_i, g_i \in \mathcal{F}, i=1,2$, the following are equivalent:
\begin{enumerate}
\item $(f_1,g_1) \sim (f_2,g_2)$

\item $\Phi(f_1,g_1) = \Phi(f_2,g_2)$.
\end{enumerate}
\end{proposition}
\proof
See Appendix \ref{equiv-proof}.
\qed

A simple example:  take $f\in\mathcal{F}$, and
define $Jf$ by $(Jf)(q,0)=-if(q,0)$, $(Jf)(0,q)=if(0,q)$ and $(Jf)(k,k')=0$ for all other
values of $k$ en $k'$. Then $(f,0)\sim (0,Jf)$ or in other words
\beq\label{rho-A}
\mathrm{CLT}-\lim_{q\to 0}\lim_{L\to\infty}\rho_{L,q}(f)
= \mathrm{CLT}-\lim_{q\to 0}\lim_{L\to\infty}A_{L,q}(Jf).
\eeq

\subsection{Dynamics of the collective Goldstone modes}\label{dynamics-1}
In this section we will derive a dynamics on the level of the macroscopic
fluctuations. This dynamics will of course be induced by the microdynamics.
Therefore we start with calculating
\begin{eqnarray}
i[H_L,\rho_{L,q}]&=&\frac{i}{2(\rho_0 V)^{1/2}}\sum_k\left[ (\epsilon_{k+q}-
\epsilon_k)a^*_{L,k+q}a_{L,k} + (\epsilon_{k-q}- \epsilon_k)a^*_{L,k-q}
a_{L,k} \right] \nonumber\\
&=& \rho_{L,q}(\tilde\epsilon),
\end{eqnarray}
with $\tilde\epsilon(k,k')=i(\epsilon_k - \epsilon_{k'})$.

And also
\begin{eqnarray}
i[H_L,A_{L,q}] &=& -\frac{1}{2}\left(\epsilon_q+(\frac{\lambda}{V}
N_L - \mu_L) \right)\left( a^*_{L,q} + a^*_{L,-q} +  a_{L,-q}+a_{L,q} \right)
\nonumber\\
&&-\frac{i\lambda}{2V}\left(a^*_{L,q} + a^*_{L,-q} - (a_{L,-q}+a_{L,q})\right)
\nonumber\\
&=&-A_{L,q}(\tilde \epsilon) -\frac{1}{2}\left(\frac{\lambda}{V}
N_L - \mu_L \right)\left( a^*_{L,q} + a^*_{L,-q} +  a_{L,-q}+a_{L,q} \right)
\nonumber\\
&&-\frac{i\lambda}{2V}\left(a^*_{L,q} + a^*_{L,-q} - (a_{L,-q}+a_{L,q})\right)
\end{eqnarray}
The second and the third term on the r.h.s. converge to zero as $L \to \infty$,
even as operators in the GNS repesentation of $\omega$, so they are of
no importance.

Remark that both $\lim_{L\to\infty}\omega(\rho_{L,q}
(\tilde\epsilon)^2)\propto \epsilon_q^2$ and $\lim_{L\to\infty}
\omega(A_{L,q} (\tilde\epsilon)^2)\propto \epsilon_q^2$, so it is
natural to define a macroscopic dynamics by
\begin{eqnarray*}
i[\tilde H, \tilde \rho]&=&\mathrm{CLT}-\lim_{q\to 0}\lim_{L\to\infty}
i\left\lbrack\frac{1}{\epsilon_q}H_L,\rho_{L,q}\right\rbrack\\
&=&\mathrm{CLT}-\lim_{q\to 0}\lim_{L\to\infty}\rho_{L,q}\left(
\frac{1}{\epsilon_q}\tilde\epsilon\right) \\
&=&\mathrm{CLT}-\lim_{q\to 0}\lim_{L\to\infty}A_{L,q}\\
&=& \tilde A,
\end{eqnarray*}
where we have used equation (\ref{rho-A}) to go from the second line to the
third.

Analogously,
\begin{eqnarray*}
i[\tilde H, \tilde A]&=&\mathrm{CLT}-\lim_{q\to 0}\lim_{L\to\infty}
i\left\lbrack\frac{1}{\epsilon_q}H_L,A_{L,q}\right\rbrack\\
&=&-\mathrm{CLT}-\lim_{q\to 0}\lim_{L\to\infty}A_{L,q}\left(
\frac{1}{\epsilon_q}\tilde\epsilon\right)\\
&=&-\mathrm{CLT}-\lim_{q\to 0}\lim_{L\to\infty}\rho_{L,q}\\
&=& -\tilde \rho,
\end{eqnarray*}
again using (\ref{rho-A}).

Hence we have found a canonical pair of observables $\tilde \rho$ and 
$\tilde A$, satisfying
\beq
[\tilde\rho,\tilde A]=i,
\eeq
which dynamically decouple from the other degrees of freedom of the system, with
the dynamics given by
\begin{eqnarray}
i[\tilde H, \tilde \rho] &=& \tilde A\\
i[\tilde H, \tilde A] &=& -\tilde \rho.
\end{eqnarray}
So $\tilde H$ is the harmonic oscillator Hamiltonian with frequency $1$:
\beq
\tilde H = \frac{1}{2}\left( \tilde\rho^2 +\tilde A^2 \right).
\eeq
The virial theorem is also satisfied, i.e.
\beq
\tilde \omega\left( \tilde\rho^2 \right) = \tilde \omega\left( \tilde A^2
 \right).
\eeq

Remark that to go from the microdynamics $H_L$ to the macrodynamics $\tilde H$
we had to rescale the Hamiltonian with $\epsilon_q^{-1}$. This should actually
be seen as a rescaling of time
\[
t \to \tilde t= \frac{t}{\epsilon_q},
\]
indicating that for small $|q|$ the typical lifetime of a (density) fluctuation
with wave length $|q|^{-1}$ is of the order $|q|^{-2}$, becoming infinite in the
limit $q\to 0$.

\section{The weakly interacting Bose gas}\label{sectionwibg}
\subsection{The model and equilibrium states}
Our second model is a model of superfluidity, i.e. with an
excitation spectrum $E_q$  linear in $|q|$ for small $q$. Such a model is
provided by \cite{19}. Its Hamiltonian is given by (we take $\nu = 3$ throughout this
section)
\begin{eqnarray}\label{wibgham}
H_L(c)&=&\sum_k \epsilon_k a^*_{L,k}a_{L,k} + \frac{1}{2}\sum_{k\not=0}v(k)
(a^*_{L,k}a^*_{L,-k} c^2 + \bar c^2 a_{L,-k}a_{L,k})\nonumber\\
&& + |c|^2 \sum_{k\not=0}
v(k)a^*_{L,k}a_{L,k} + \frac{v(0)}{2V}N_L^2 - \mu_L N_L,
\end{eqnarray}
supplemented with
\beq
c=\lim_{L\to\infty}\omega_L(V^{-1/2}a_{L,0}),
\eeq
where $\omega_L$ is the Gibbs state at some inverse temperature $\beta$ corresponding 
to (\ref{wibgham}).

This Hamiltonian is in fact the original Bogoliubov Hamiltonian for a weakly
interacting Bose gas, with an extra term $\frac{v(0)}{2V}N_L^2$ that ensures the
superstability of the model.

Again we take the thermodynamic limit under the constraint
\[
\lim_{L\to\infty}\frac{1}{V}\omega_L(N_L)=\rho.
\]
It is proved in \cite{19} that there exist solutions $\omega_\beta=\lim_{L\to\infty}
\omega_L$, for $\beta$ and $\rho$ large enough, such that
\[
\lim_L\omega_\beta(V^{-1/2}a_{L,0})=c\not=0,
\]
and we will restrict ourself to these solutions. 

To describe these equilibrium states we need a Bogoliubov transformation of the
operators $a^\sharp_{L,k}$ into new creation and annihilation operators 
$b^\sharp_{L,k}$:
\begin{eqnarray}
a_{L,k} &=& b_{L,k} \cosh \alpha_k + b^*_{L,-k} \sinh \alpha_k \label{bog1}\\
a_{L,-k} &=& b_{L,-k} \cosh \alpha_k + b^*_{L,k} \sinh \alpha_k,\label{bog2}
\end{eqnarray}
where
\beq\label{tanh}
\tanh 2\alpha_k = -\frac{|c|^2 v(k)}{\epsilon_k + |c|^2 v(k)}.
\eeq

The limit Gibbs state has the same form as in the imperfect Bose gas:
for all local observables $A$,
\beq
\omega_\beta(A) = \frac{1}{2\pi}\int_0^{2\pi} \omega_\beta^\alpha(A) d\alpha,
\eeq
with
\beq\label{state2}
\omega_\beta^\alpha\left( e^{i(b(f) + b^*(f))}  \right)
= \exp\left[ -\frac{1}{2}(f,K'f) + 2i|c \hat f(0)| \cos \alpha \right]
\eeq
and
\beq
\left( \widehat{K'f} \right) (k) = \frac{1}{2}\coth\left(\frac{\beta E_k}{2}
\right) \hat f(k),
\eeq
with $b^\sharp(f)$ the corresponding Bogoliubov transformation of $a^\sharp(f)$
and
\beq\label{E}
E_k = \sqrt{(\epsilon_k + |c|^2v(k))^2 - (|c|^2v(k))^2}
= \sqrt{\epsilon_k (\epsilon_k + 2|c|^2v(k))}.
\eeq
This is the famous Bogoliubov spectrum which for small $k$ behaves like 
\[
E_k \simeq \left( \frac{|c|^2v(0)}{m} \right)^{1/2}|k|.
\]

The states $\omega_\beta^\alpha (\alpha\in [0,2\pi] )$ are the quasi-free extremal
equilibrium state components of $\omega_\beta$ with the property that
\beq
\lim_{L\to\infty}\omega_\beta^\alpha \left(\frac{a^*_{L,0}}{V^{1/2}} \right) =
|c|e^{i\alpha},
\eeq
and as operators in the GNS-representation of $\omega_\beta^\alpha$, one has
also
\beq\label{a02}
\lim_{L\to\infty}\frac{a^*_{L,0}}{V^{1/2}} =
|c|e^{i\alpha}.
\eeq

For more details we refer to \cite{19}.

As in the imperfect Bose gas, we want to study the density and order parameter
fluctuations, $F_{L,q}(N)$ and $F_{L,q}(A)$. However here, 
the Hamiltonian $H_L(c)$
(\ref{wibgham}) is a truncation of the full Hamiltonian (\ref{genH}), and due to this
truncation, $H_L(c)$ is no longer gauge invariant, i.e.
\beq\label{[H,N]}
[H_L(c),N_L]\not= 0.
\eeq
The invariance which is left is
\beq\label{[H,N0]}
[H_L(c),N_{L,0}]=0,
\eeq
with $N_{L,0}=a^*_{L,0}a_{L,0}$
This means that the spontaneously broken symmetry accompanying the phase transition
from $c=0$ to $c\not= 0$ is not the gauge symmetry generated by $N_L$, but the symmetry
generated by $N_{L,0}$.

One example of the implications of this is the following.
In the physics literature, the quantity $\lim_{L\to\infty}<F_{L,q}(N)
F_{L,-q}(N)>$ is known as the static structure function, usually denoted $S(q)$.
It has been known for a long time, both theoretically and experimentally, that
at zero temperature this
function behaves linearly in $q$ for small $q$: $S(q)\propto |q|$
(see e.g. \cite{20}). This linear behaviour is essentially due to the fact that $[U_L,
F_{L,q}(N)]=0$ so that 
\[
<[F_{L,q}(N),[H_L,F_{L,-q}(N)]]> \propto |q|^2.
\]
However in our model 
\[
[H_L(c),N_L]\not= 0,
\]
and very much related to this,
\[
[U_L(c),F_{L,q}(N)]\not=0,
\]
and indeed it is easy to calculate that here
$\lim_{q\to 0}S(q)=\mathrm{const}\not=0$.

Because of (\ref{[H,N]}) and (\ref{[H,N0]}) we expect that this unphysical behaviour
is remedied when we replace the total density fluctuations $F_{L,q}(N)$ by condensate
density fluctuations $F_{L,q}(N_0)$. However since $N_{L,0}$ can not be written as
the integral over some condensate density, it is impossible to define $F_{L,q}(N_0)$
as a usual fluctuation operator. What we want to show now is that it is possible to
find a fluctuation operator $F_{L,q}(N_0)$ which behaves mathematically 
like one expects for a
fluctuation operator of the generator of a spontaneously broken symmetry (i.e. we
will derive a similar structure as in the imperfect Bose gas)
and moreover gives the correct physical behaviour (like e.g.
\[
\lim_{L\to\infty}<F_{L,q}(N_0)F_{L,-q}(N_0)>\propto |q|
\]
 for small $q$).

In momentum space we have
\beq
F_{L,q}(N) = \frac{1}{V^{1/2}}\sum_{k}a^*_{L,k+q}a_{L,k}.
\eeq
This consists of two parts
\beq
F_{L,q}(N) = \frac{1}{V^{1/2}}\left(a^*_{L,q}a_{L,0} + a^*_{L,0}a_{L,-q} 
\right) + \frac{1}{V^{1/2}}\sum_{k\not= 0, k+q\not=0}a^*_{L,k+q}a_{L,k}.
\eeq
The first part
\[
\frac{1}{V^{1/2}}\left(a^*_{L,q}a_{L,0} + a^*_{L,0}a_{L,-q} 
\right)
\]
is the part of $F_{L,q}(N)$
which contains the ground state operators $a^\sharp_{L,0}$.
It is clearly the fluctuation of the zero-mode particle density, fluctuating to a
fixed mode and back to zero. The other part is the fluctuation of the excited
modes among each other. Therefore it is natural to define
\beq
F_{L,q}(N_0)=\frac{1}{V^{1/2}}\left(a^*_{L,q}a_{L,0} + a^*_{L,0}a_{L,-q}
\right).
\eeq

The truncation of $F_{L,q}(N)$ to $F_{L,q}(N_0)$ reminds very much the spirit behind 
the truncation which
led to the Bogoliubov approximation of the full Hamiltonian. We can even show how
closely those two are related. Take the interaction part $U_L$ of the full Hamiltonian
(\ref{genH}) and write it as in (\ref{U}):
\[
U_L=\frac{1}{2}\sum_{k\not=0}v(k)F_{L,k}(N)F_{L,-k}(N) + 
\frac{v(0)}{2V}N_L^2-\frac{1}{2}\phi(0)N_L.
\]
Truncate this expression by truncating the operators $F_{L,k}(N)$ to $F_{L,k}(N_0)$
as described above, then:
\[
U_L=\frac{1}{2}\sum_{k\not=0}v(k)F_{L,k}(N_0)F_{L,-k}(N_0) + 
\frac{v(0)}{2V}N_L^2-\frac{1}{2}\phi(0)N_L.
\]
Write out:
\begin{eqnarray*}
&&\frac{1}{2}\sum_{k\not=0}v(k)F_{L,k}(N_0)F_{L,-k}(N_0)\\
&&=\frac{1}{2V}\sum_{k\not=0}v(k)\left(
a_{L,0}a^*_{L,0}a^*_{L,k}a_{L,k} + a^*_{L,0}a_{L,0}a_{L,-k}a^*_{L-k} \right.\\
&&\left. \;\;\;+ a_{L,0}a_{L,0}a^*_{L,k}a^*_{L,-k} + a^*_{L,0}a^*_{L,0}a_{L,-k}a_{L,k}
\right)\\
&&=\frac{1}{2V}\sum_{k\not=0}v(k)\left( (a_{L,0}a^*_{L,0}+
a^*_{L,0}a_{L,0})a^*_{L,k}a_{L,k} + a_{L,0}a_{L,0}a^*_{L,k}a^*_{L,-k}
\right.\\
&&\left. \;\;\;+ a^*_{L,0}a^*_{L,0}a_{L,-k}a_{L,k}
\right) + \frac{\phi(0)}{2}N_{L,0}
\end{eqnarray*}
As in \cite{19}, replace the operators $\frac{a^\sharp_{L,0}}{V^{1/2}}$ by
complex numbers 
$|c|e^{\pm i\alpha}$
in this part of the interaction and preserve them as operators in the
remaining terms, then 
\begin{eqnarray}
U_L &=&  \frac{1}{2}\sum_
{k\not= 0}v(k)|c|^2\left(e^{-2i\alpha}a^*_{L,k}a^*_{L,-k}
+e^{2i\alpha}a_{L,-k}a_{L,k} \right)
\nonumber \\
&&+|c|^2\sum_{k\not= 0}v(k)a^*_{L,k}a_{L,k}
+\frac{v(0)}{2V}N_L^2- \frac{1}{2}\phi(0)N_L +\frac{\phi(0)}{2}
 |c|^2 V.
\end{eqnarray}
Apart from the term $\frac{1}{2}\phi(0)N_L$, which only leads to a shift in the
chemical potential, and
an unimportant constant $\frac{\phi(0)}{2} |c|^2 V$  this is exactly the
Hamiltonian (\ref{wibgham}).

\subsection{Collective Goldstone modes}
We consider again our system to be in one of the extremal equilibrium states
$\omega_\beta^\alpha$, and without loss of generality we take $\alpha = 0$
(i.e. $c$ real), and
denote this state by $\omega_\beta$. As before, let
\begin{eqnarray*}
F_{L,q}(N_0) &=& \frac{1}{V^{1/2}}\left(a^*_{L,q}a_{L,0} + a^*_{L,0}a_{L,-q}
\right)\\
F_{L,q}(A) &=& \frac{i}{\sqrt{2}V^{1/2}}\int_{\Lambda}(a^*(x)-a(x))e^{iq.x}dx\\
&=& \frac{i}{\sqrt{2}}(a^*_{L,q} - a_{L,-q}),
\end{eqnarray*}
and for ease of notation:
\[
\rho^0_{L,q}=\frac{1}{\sqrt{2c^2}}F_{L,q}(N_0), \; \;\;\;A_{L,q}=F_{L,q}(A).
\]
These operators still satisfy the correct commutation relation
\[
[\rho^0_{L,q},A_{L,-q}]=\frac{i}{2\sqrt{c^2 V}}(a^*_{L,0} + a_{L,0}),
\]
and by (\ref{a02}):
\beq
\lim_{L\to\infty}[\rho^0_{L,q},A_{L,-q}]=i.
\eeq
More generally
\beq
\lim_{L\to\infty}[\rho^0_{L,q},A_{L,-q'}]=i\delta_{q,q'}.
\eeq

\begin{proposition}We have for $q,q'\not=0$
\begin{enumerate}
\item
\[
\lim_{L\to\infty}\omega_\beta(\rho^0_{L,q}\rho^0_{L,-q'}) = \delta_{q,q'}
\frac{\epsilon_q}{2E_q}\coth\left( \frac{\beta E_q}{2}
\right)
\]

\item
\[
\lim_{L\to\infty}\omega_\beta(A_{L,q}A_{L,-q'}) = \delta_{q,q'}
\frac{E_q}{2\epsilon_q}\coth\left( \frac{\beta E_q}{2}
\right).
\]
\end{enumerate}
\end{proposition}
\proof
This is an easy calculation using the Bogoliubov transformation (\ref{bog1}),
(\ref{bog2}), the explicit form of the state (\ref{state2}), property (\ref{a02})
 and the fact that
\begin{eqnarray*}
\cosh 2\alpha_q&=&\frac{\epsilon_q+c^2v(q)}{E_q}\\
\sinh 2\alpha_q&=&-\frac{c^2v(q)}{E_q},
\end{eqnarray*}
so that
\begin{eqnarray}
(\cosh \alpha_q + \sinh\alpha_q)^2 &=&\cosh 2\alpha_q + \sinh2\alpha_q
\nonumber\\
&=&\frac{\epsilon_q}{E_q} \nonumber\\
&=&\sqrt{  \frac{\epsilon_q}{\epsilon_q+2c^2v(q)} }\\
(\cosh \alpha_q - \sinh\alpha_q)^2 &=&\cosh 2\alpha_q - \sinh2\alpha_q
\nonumber\\
&=& \frac{\epsilon_q+2c^2v(q)}{E_q} \nonumber\\
&=&\frac{E_q}{\epsilon_q}\nonumber\\
&=&\sqrt{  \frac{\epsilon_q+2c^2v(q)}{\epsilon_k} }.
\end{eqnarray}
This then gives the result via
\begin{eqnarray*}
\lim_{L\to\infty}\omega_\beta(\rho^0_{L,q}\rho^0_{L,-q'}) &=& \delta_{q,q'}\frac{1}{2}
\lim_{L\to\infty}\omega_\beta\left( (a^*_{L,q}+a_{L,-q}) (a^*_{L,-q}+a_{L,q})\right)\\
&=&\delta_{q,q'}\frac{1}{2}(\cosh \alpha_q + \sinh\alpha_q)^2 \times\\
&&\underbrace{\lim_{L\to\infty}\omega_\beta\left( (b^*_{L,q}+b_{L,-q}) 
(b^*_{L,-q}+b_{L,q})\right)}_{\coth\left( \frac{\beta E_q}{2}
\right)},
\end{eqnarray*}
\qed

For $\beta < \infty$ the small $q$-behaviour of these variances is
\begin{eqnarray*}
\lim_{L\to\infty}\omega_\beta(\rho^0_{L,q}\rho^0_{L,-q})&\simeq& \mathrm{const}
\not= 0\\
\lim_{L\to\infty}\omega_\beta(A_{L,q}A_{L,-q}) &\simeq& \mathrm{const}\times |q|^{-2}. 
\end{eqnarray*}
So again we have the phenomenon that at non-zero temperature it is impossible to
do a renormalisation of $\rho^0_{L,q}$ and $A_{L,q}$ which gives both 
a meaningful  $q\to 0$ limit for the variances and preserves a non-trivial
commutation relation.

Therefore we will restrict ourself from now on to the ground state (denoted
$\omega$). Contrary to the imperfect Bose gas, there remains a
non-trivial $q$-dependence in the ground state. This is because even at zero
temperature not all particles condense into the ground state.
We have for small $q$
\begin{eqnarray*}
\lim_{L\to\infty}\omega(\rho^0_{L,q}\rho^0_{L,-q})&\propto& |q|\\
\lim_{L\to\infty}\omega_\beta(A_{L,q}A_{L,-q}) &\propto& |q|^{-1}. 
\end{eqnarray*}
Remark that for the condensate density fluctuations this is the above mentioned
linear behaviour.

We now redefine $\rho^0_{L,q}$ and $A_{L,q}$ to be self-adjoint and renormalised
in $q$, i.e.
\begin{eqnarray}
\rho^0_{L,q} &=& \frac{1}{2(c^2|q|V)^{1/2}}\left\lbrack 
(a^*_{L,q}+a^*_{L,-q})a_{L,0} + a^*_{L,0}(a_{L,q}+a_{L,-q})\right\rbrack\\
A_{L,q} &=& i\frac{|q|^{1/2}}{2}\left\lbrack 
a^*_{L,q}+a^*_{L,-q} - (a_{L,q}+a_{L,-q})\right\rbrack.
\end{eqnarray}
Hence we have still
\beq
\lim_{L\to\infty}[\rho^0_{L,q},A_{L,q'}]=i\delta_{q,q'}.
\eeq

It is already interesting at this stage to remark that
\begin{eqnarray}
\lim_{q\to 0}\lim_{L\to\infty}\omega_\beta\left((\rho^0_{L,q})^2\right)
&=& \frac{1}{2\Omega}\\
\lim_{q\to 0}\lim_{L\to\infty}\omega_\beta\left(A_{L,q}^2\right) &=&
\frac{\Omega}{2},
\end{eqnarray}
with
\beq\label{Omega}
\Omega =\lim_{q\to 0}\frac{E_q |q|}{\epsilon_q}= (4mc^2v(0))^{1/2}.
\eeq
And hence
\beq
\Omega^2\lim_{q\to 0}\lim_{L\to\infty}\omega\left((\rho^0_{L,q})^2\right)
=\lim_{q\to 0}\lim_{L\to\infty}\omega\left(A_{L,q}^2\right).
\eeq

Again the rest of this section is devoted to the rigorous mathematical treatment of the
existence of the fluctuation operators. Again at first reading, the reader can immediately
proceed to section \ref{dynamics-2}.

With all definitions and notations as above for the imperfect Bose gas, we define
again for $f,g \in\mathcal{F}$
\begin{eqnarray}
\rho_{L,q}^0(f)&=&\frac{1}{2c V^{1/2}}\left\lbrack f(q,0)
(a^*_{L,q}+a^*_{L,-q})a_{L,0} \right.\nonumber\\
&&\left.+ f(0,q) a^*_{L,0}(a_{L,q}+a_{L,-q})
 \right\rbrack\\
A_{L,q}(g) &=& \frac{i}{2}\left\lbrack g(q,0)
(a^*_{L,q}+a^*_{L,-q}) -g(0,q) (a_{L,q}+a_{L,-q})\right\rbrack,
\end{eqnarray}
and
\beq
F_{L,q}(f,g) = \rho_{L,q}^0(f) + A_{L,q}(g).
\eeq

Now we take immediately the double limit $\lim_{q\to 0}\lim_{L\to\infty}$ rather than
the two limits separately as for the imperfect Bose gas.
\begin{proposition}
For $f,g \in\mathcal{F}$,
\[
\lim_{q\to 0}\lim_{L\to\infty}\omega\left( F_{L,q}(f,g)^2 \right)
\]
\beq\label{F^2}
=\lim_{q\to 0}\left\{ \frac{\epsilon_q + c^2v(q)}{2E_q}\Bigl| f(q,0)+ig(q,0) \Bigr|^2
- \frac{c^2v(q)}{2E_q}\Re\left\lbrack \Bigl(f(q,0)+ig(q,0)\Bigr)^2 \right\rbrack
  \right\}.
\eeq
\end{proposition}
\proof
An explicit calculation shows that
\[
\lim_{L\to\infty}\omega\left(\rho_{L,q}^0(f)^2\right) =
\frac{\epsilon_q + c^2v(q)}{2E_q} |f(q,0)|^2 - \frac{c^2v(q)}{4E_q}
[f(q,0)^2+f(0,q)^2],
\]
\[
\lim_{L\to\infty}\omega\left(A_{L,q}(g)^2\right) =
\frac{\epsilon_q + c^2v(q)}{2E_q} |g(q,0)|^2 + \frac{c^2v(q)}{4E_q}
[g(q,0)^2+g(0,q)^2]
\]
and
\begin{eqnarray*}
&&\omega\left( \rho_{L,q}^0(f)A_{L,q}(g) + A_{L,q}(g)\rho_{L,q}^0(f)\right)\\
&=&\frac{\epsilon_q + c^2v(q)}{E_q}\Im [f(q,0)g(0,q)] +
\frac{c^2v(q)}{E_q}\Im [f(q,0)g(q,0)],
\end{eqnarray*}
which together give (\ref{F^2}).
\qed
From now on we restrict to those pairs of functions $(f,g)$ for which (\ref{F^2})
remains finite. This means if $f$ is real, $|f(q,0)|$ should not diverge faster than 
$|q|^{-1/2}$ and if $f$ is imaginary, $|f(q,0)|$ should go to zero, at least like
$|q|^{1/2}$. A real $g$ should satisfy the same condition as an imaginary $f$ and vice
versa. Alternatively, we could say we restrict $\mathcal{F}$ to those functions $f$ which
satisfy the above conditions and then look at pairs $(f,Jg)$, with 
$f,g$ in (the restricted) $\mathcal{F}$ and
$Jg$ defined as $(Jg)(q,0)=-ig(q,0)$, $(Jg)(0,q)=ig(0,q)$.

\begin{proposition}[A BCH-formula]\label{BCH-prop2}
Let $(f_i,g_i) \in (\mathcal{F},J\mathcal{F}), i=1,2$, then
\[
\lim_{L\to\infty}\left\| e^{iF_{L,q}(f_1,g_1)}e^{iF_{L,q}(f_2,g_2)} -
e^{i(F_{L,q}(f_1,g_1) + F_{L,q}(f_2,g_2))}
e^{-\frac{1}{2}[F_{L,q}(f_1,g_1),F_{L,q}(f_2,g_2)]} \right\|_\omega
\]
\beq\label{BCH2}
 = 0.
\eeq
\end{proposition}
\newpage
\proof
See Appendix \ref{BCH-append}.
\qed

On $\mathcal{V}$, the space of complex linear combinations of elements from
$(\mathcal{F},J\mathcal{F})$, $J\mathcal{F}$ in the above defined sense,
 define a positive sesquilinear form $<\cdot|\cdot>$ by
\[
<f_1,g_1|f_2,g_2>=\lim_{q\to 0}\lim_{L\to\infty}\omega
\Bigl( F_{L,q}(f_1,g_1)^*F_{L,q}(f_2,g_2) \Bigr),
\]
and extension to $\mathcal{V}$ by linearity.
This can be calculated in the same way as (\ref{F^2}) is calculated to give the following
expression
\[
\lim_{L\to\infty}\omega\Bigl( F_{L,q}(f_1,g_1)F_{L,q}(f_2,g_2) \Bigr)
\]
\begin{eqnarray*}
=&&\frac{\epsilon_q + c^2v(q)}{2E_q}\Re\left(\overline{[f_1+ig_1]}[f_2+ig_2]\right)
-\frac{c^2v(q)}{2E_q}\Re\Bigl([f_1+ig_1][f_2+ig_2]\Bigr)\\
&&+i\Im\left(\overline{[f_1+ig_1]}[f_2+ig_2]\right).
\end{eqnarray*}
where we adopted the short hand notation $f_i=f_i(q,0)$, $\bar f_i=f_i(0,q)$, etc.

In practice all operators we
use, have either $f=0$ or $g=0$, and $f$ and $g$ will always be
either real or imaginary. It is easily seen that this leads to a considerable
simplification of the above formula's. 

Separating the real and the imaginary part of the restriction of $<\cdot|\cdot>$ to the
real subspace $(\mathcal{F},J\mathcal{F})$ of $\mathcal{V}$,
\beq
<f_1,g_1|f_2,g_2>=s(f_1,g_1|f_2,g_2)+\frac{i}{2}\sigma(f_1,g_1|f_2,g_2),
\eeq
defines a real bilinear positive symmetric form $s$ and a symplectic form $\sigma$.

Remark that, as it should, this $\sigma$ is the same as the one for the imperfect Bose gas
since it also satisfies
\beq
\lim_{q\to 0}\lim_{L\to\infty}[F_{L,q}(f_1,g_1),F_{L,q}(f_2,g_2)] = i
\sigma(f_1,g_1|f_2,g_2),
\eeq
where the limit is taken in the GNS-representation of the equilibrium state $\omega$. 
This commutator is of
course independent of the given model since in both models we have
\[
V^{-1/2}a^\sharp_{L,0}\to c.
\]

\begin{proposition}[Central Limit Theorem]
For $(f,g) \in (\mathcal{F},J\mathcal{F})$, $t\in\R$
\beq\label{clt2}
\lim_{q\to 0}\lim_{L\to\infty}\omega\left( e^{itF_{L,q}(f,g)} \right)
=e^{-\frac{t^2}{2}s\left(f,g|f,g\right)}.
\eeq
\end{proposition}
\proof
Using (\ref{a02}) we get 
\[
\lim_{q\to 0}\lim_{L\to\infty}\omega\left( e^{itF_{L,q}(f,g)} \right)=
\]
\beq\label{rho00} 
\lim_{q\to 0}\lim_{L\to\infty}\omega\left( e^{\frac{it}{2}\left\{ \left[f(q,0)
+ig(q,0) \right]\left(a^*_{L,q}+a^*_{L,-q}\right) 
+ \overline{\left[f(q,0)+ig(q,0) \right]} \left(a_{L,q}+a_{L,-q}\right) \right\}}
\right).
\eeq 
With the explicit expression of the state (\ref{state2}), the r.h.s. of (\ref{rho00})
can easily be computed and we get (\ref{clt2}).
\qed

\begin{proposition}[Reconstruction Theorem]
The linear functional
\beq
\tilde \omega\left( W(f,g) \right)= e^{-\frac{1}{2}s(f,g|f,g)}
\eeq
defined on the algebra $CCR((\mathcal{F},J\mathcal{F}),\sigma)$, is a quasi free state.

More explicitly, we have for all $(f_i,g_i) \in (\mathcal{F},J\mathcal{F}), 
i=1,\ldots,n$,
\beq\label{clt-expl3}
\lim_{q\to 0}\lim_{L\to\infty}\omega\left(  e^{iF_{L,q}(f_1,g_1)}\cdots
 e^{iF_{L,q}(f_n,g_n)}\right) = \tilde \omega \left( W(f_1,g_1)
\cdots W(f_n,g_n) \right).
\eeq

The state $\tilde \omega$ is regular and hence for every $(f,g)$ there
exists a self-adjoint Bosonic field $\Phi(f,g)$ in the GNS representation
$(\mathcal{H}_{\tilde \omega},\pi_{\tilde \omega},
\Omega_{\tilde \omega})$ such that
\beq
\pi_{\tilde \omega}\left( W(f,g) \right) = e^{i\Phi(f,g)}.
\eeq

This implies that in this sense of the central limit (\ref{clt-expl3}), 
the local fluctuations converge to the Bosonic fields associated with 
$CCR((\mathcal{F},J\mathcal{F}),\sigma)$:
\beq
\Phi(f,g)=\mathrm{CLT}-\lim_{q\to 0}\lim_{L\to\infty}
F_{L,q}(f,g).
\eeq
\end{proposition}
\proof
See the proof of Proposition \ref{recthm} and Proposition \ref{recthm2}.
\qed

Specifying to our original operators:
\begin{eqnarray}
\tilde \rho^0 &=& \Phi(|q|^{-1/2},0) = \mathrm{CLT}-\lim_{q\to 0}\lim_{L\to\infty} 
\rho^0_{L,q} \\
\tilde A &=& \Phi(0,|q|^{1/2}) = \mathrm{CLT}-\lim_{q\to 0}\lim_{L\to\infty} A_{L,q},
\end{eqnarray}
adopting the notation $(f,g)=(f(q,0),g(q,0))$.

The equivalence property also remains true in this case of course, i.e.
\begin{eqnarray*}
\Phi(f_1,g_1) = \Phi(f_2,g_2) &\iff& (f_1,g_1) \sim (f_2,g_2) \\
&\iff& \lim_{q\to 0}\lim_{L\to\infty} 
\omega\left( F_{L,q}(f_1-f_2,g_1-g_2)^2 \right) =0,
\end{eqnarray*}
in particular
\beq
\mathrm{CLT}-\lim_{q\to 0}\lim_{L\to\infty} \rho^0_{L,q}(f) =
\mathrm{CLT}-\lim_{q\to 0}\lim_{L\to\infty} A_{L,q}(Jf).
\eeq

\subsection{Dynamics of the collective Goldstone modes}\label{dynamics-2}
We will now show that also for the weakly interacting Bose gas the fluctuations of the
generator  and of the order parameter of the SSB decouple dynamically from the other
degrees of freedom of the system, and form a harmonic oscillator system.

At the local level, we have
\begin{eqnarray}
&&i[H_L(c),\rho^0_{L,q}]\nonumber\\
&=& \frac{i\epsilon_q }{2(c^2|q|V)^{1/2}}\left\lbrack 
(a^*_{L,q}+a^*_{L,-q})a_{L,0} - a^*_{L,0}(a_{L,q}+a_{L,-q})\right\rbrack\nonumber\\
&&+\frac{ic^2v(q)}{2(c^2|q|V)^{1/2}}\left\lbrack 
(a^*_{L,q}+a^*_{L,-q})(a_{L,0}-a^*_{L,0}) + (a_{L,0}-a^*_{L,0})
(a_{L,q}+a_{L,-q})\right\rbrack\nonumber\\
&=&\rho^0_{L,q}\left( \frac{i\epsilon_q}{|q|^{1/2}} \right)
+\frac{ic^2v(q)}{2(c^2|q|V)^{1/2}}\left\lbrack 
(a^*_{L,q}+a^*_{L,-q})(a_{L,0}-a^*_{L,0}) \right.\nonumber\\
&&\left.+ (a_{L,0}-a^*_{L,0})
(a_{L,q}+a_{L,-q})\right\rbrack.
\end{eqnarray}

Since the operators $V^{-1/2}(a_{L,0}-a^*_{L,0})$ converge to 0 in the GNS representation
of $\omega$, we only have to take into account the term 
$\rho^0_{L,q}\left( \frac{i\epsilon_q}{|q|^{1/2}} \right)$. 
This is what we mentioned earlier, the interaction
part of $H_L(c)$ commutes with the condensate density fluctuations as one expects
physically, but only in the thermodynamic limit.

Also:
\begin{eqnarray}
&&-i[H_L(c),A_{L,q}]\nonumber\\
&=&\frac{(\epsilon_q + 2c^2v(q))|q|^{1/2}}{2}\left\lbrack 
a^*_{L,q}+a^*_{L,-q} + a_{L,q}+a_{L,-q}\right\rbrack \nonumber\\
&=& A_{L,q}\left(-i(\epsilon_q+2c^2v(q))|q|^{1/2} \right).
\end{eqnarray}

As in the imperfect Bose gas, we define a macroscopic dynamics by first renormalizing
the Hamiltonian with its spectrum,  and then taking the limit. 
$\mathrm{CLT}-\lim_{q\to 0}\lim_{L\to\infty}$ of the commutators.
\begin{eqnarray}
i[\tilde H, \tilde \rho^0] 
&=& \mathrm{CLT}-\lim_{q\to 0}\lim_{L\to\infty} i\left\lbrack \frac{1}{E_q}H_L(c),
\rho^0_{L,q}\right\rbrack\nonumber\\
&=&\mathrm{CLT}-\lim_{q\to 0}\lim_{L\to\infty}
\rho^0_{L,q}\left(\frac{i\epsilon_q}{E_q |q|^{1/2}}  \right)\nonumber\\
&=&\mathrm{CLT}-\lim_{q\to 0}\lim_{L\to\infty}A_{L,q}\left(
\frac{\epsilon_q}{E_q |q|}|q|^{1/2}\right)
\nonumber\\
&=&\frac{1}{\Omega}\tilde A
\end{eqnarray}
(where we used the equivalence relation). Remember that $A_{L,q}=A_{L,q}(|q|^{1/2})$ 
and that $\Omega$ was defined in (\ref{Omega}),
\[
\Omega = (4mc^2v(0))^{1/2}.
\]
\begin{eqnarray}
-i[\tilde H,\tilde A]
&=&\mathrm{CLT}-\lim_{q\to 0}\lim_{L\to\infty} -i\left\lbrack \frac{1}{E_q}H_L(c),
A_{L,q}\right\rbrack\nonumber\\
&=&\mathrm{CLT}-\lim_{q\to 0}\lim_{L\to\infty}
A_{L,q}\left(\frac{-i(\epsilon_q+2c^2v(q))|q|^{1/2}}{E_q} \right)\nonumber\\
&=&\mathrm{CLT}-\lim_{q\to 0}\lim_{L\to\infty}\rho^0_{L,q}\left(
\frac{E_q|q|}{\epsilon_q}\frac{1}{|q|^{1/2}} \right)\nonumber\\
&=&\Omega \tilde \rho^0.
\end{eqnarray}
Again remember $\rho^0_{L,q}=\rho^0_{L,q}(|q|^{-1/2})$.

Hence we have found a canonical pair $\tilde \rho^0$ and $\tilde A$, satisfying
\beq
[\tilde \rho^0,\tilde A]=i,
\eeq
which dynamically decouple from the other degrees of freedom of the system, with the
dynamics given by
\begin{eqnarray}
i[\tilde H,\tilde \rho^0]=\frac{1}{\Omega}\tilde A\\
-i[\tilde H,\tilde A]=\Omega \tilde \rho^0.
\end{eqnarray}
The solution of these equations is the harmonic oscillator with energy $\Omega$:
\beq
\tilde H = \frac{1}{2\Omega}\left(\Omega^2  (\tilde \rho^0)^2 + \tilde A^2 \right).
\eeq

The state $\tilde\omega$ is an equilibrium ground state 
for the Hamiltonian $\tilde H$ satisfying the following virial theorem
\beq
\Omega^2\tilde\omega\left( (\tilde \rho^0)^2 \right) =
\omega\left( \tilde A^2 \right).
\eeq

We classified the programme of Anderson \cite{13,14} about the construction of the
canonical Goldstone coordinates.

Remark that also in this case we have done a rescaling of time
\[
t\to \tilde t=\frac{t}{E_q}
\]
when going from the microdynamics $H_L(c)$ to the macrodynamics $\tilde H$,
indicating again that the typical lifetime of the fluctuations becomes infinite as $q
\to 0$, this time with a rate $|q|^{-1}$, as a consequence of the fact that $E_q \propto
|q|$ for small $q$.

\section*{Acknowledgements}
The authors thank V.A. Zagrebnov for pointing out the references \cite{13,14} and for
discussions prior to this work.

\appendix
\section*{Appendices}

\section{Proof of Proposition \ref{BCH-prop} and \ref{BCH-prop2}}
\label{BCH-append}
In this proof we make use of the following general formula
\beq\label{dyson}
e^{it(x+y)}=e^{itx}+\int_0^t ds e^{isx}iye^{i(t-s)(x+y)},
\eeq
where $x,y$ are self-adjoint operators. From this formula, one deduces useful
properties like for example
\beq\label{dys-com1}
\lbrack e^{ix},z \rbrack = i\int_0^1 dt e^{itx}[x,z]e^{i(1-t)x},
\eeq
\beq\label{dys-com2}
\| [e^{ix},z] \| \leq \| [x,z] \|,
\eeq
where $x,y$ are self-adjoint, $z$ arbitrary and $\|.\|$ is some norm.

Denote $F_i=F_{L,q}(f_i,g_i)$, $i=1,2$. We need to prove
\[
\lim_{L\to\infty}\| e^{iF_1}e^{iF_2}-e^{i(F_1+F_2)}e^{-\frac{1}{2}[F_1,F_2]} \|_\omega=0.
\]
One has
\begin{eqnarray*}
&&\| e^{iF_1}e^{iF_2}-e^{i(F_1+F_2)}e^{-\frac{1}{2}[F_1,F_2]} \|_\omega\\
&=& \omega\left(\{e^{iF_1}e^{iF_2}-e^{i(F_1+F_2)}e^{-\frac{1}{2}[F_1,F_2]}\}^*
  \{e^{iF_1}e^{iF_2}-e^{i(F_1+F_2)}e^{-\frac{1}{2}[F_1,F_2]}\}\right)\\
&=& 2 - \omega\left(e^{-iF_2}e^{-iF_1}e^{i(F_1+F_2)}e^{-\frac{1}{2}[F_1,F_2]}\right)
-  \omega\left(e^{\frac{1}{2}[F_1,F_2]}e^{-i(F_1+F_2)}e^{iF_1}e^{iF_2} \right).
\end{eqnarray*}
Hence it is sufficient to show
\[
\lim_{L\to\infty}\omega\left(e^{-iF_2}e^{-iF_1}e^{i(F_1+F_2)}
e^{-\frac{1}{2}[F_1,F_2]}\right)=1.
\]
Define a function $f(t)$ by
\[
f(t)=\omega\left(e^{-iF_2}e^{-itF_1}e^{i(tF_1+F_2)}
e^{-\frac{t}{2}[F_1,F_2]}\right)-1.
\]
We have to show 
\[
\lim_{L\to\infty}|f(1)|=0.
\]
By Taylor's theorem, there exists some $t\in [0,1]$ such that
\[
f(1)=f(0)+f'(t)=f'(t),
\]
since $f(0)=0$. Let us calculate $f'(t)$ using (\ref{dyson}),
\[
\frac{d}{dt}e^{i(tF_1+F_2)}=i\int_0^1 ds e^{is(tF_1+F_2)}F_1e^{i(1-s)(tF_1+F_2)}.
\]
Hence
\begin{eqnarray*}
f'(t) &=& \omega\left( -ie^{-iF_2}e^{-itF_1}F_1e^{i(tF_1+F_2)}e^{-\frac{t}{2}[F_1,F_2]}
\right. \\
&&+ie^{-iF_2}e^{-itF_1}\int_0^1 ds e^{is(tF_1+F_2)}F_1e^{i(1-s)(tF_1+F_2)}
e^{-\frac{t}{2}[F_1,F_2]}\\
&& \left. -\frac{1}{2}e^{-iF_2}e^{-itF_1}e^{i(tF_1+F_2)}[F_1,F_2]e^{-\frac{t}{2}[F_1,F_2]}
\right)\\
&=& \omega\left( -ie^{-iF_2}e^{-itF_1}F_1e^{i(tF_1+F_2)}e^{-\frac{t}{2}[F_1,F_2]}
\right. \\
&&+ie^{-iF_2}e^{-itF_1}F_1e^{i(tF_1+F_2)}e^{-\frac{t}{2}[F_1,F_2]}\\
&&+ie^{-iF_2}e^{-itF_1}\int_0^1 ds [e^{is(tF_1+F_2)},F_1]e^{i(1-s)(tF_1+F_2)}
e^{-\frac{t}{2}[F_1,F_2]}\\
&& \left. -\frac{1}{2}e^{-iF_2}e^{-itF_1}e^{i(tF_1+F_2)}[F_1,F_2]e^{-\frac{t}{2}[F_1,F_2]}
\right)\\
&=&\omega\left(ie^{-iF_2}e^{-itF_1}\int_0^1 ds\left\{[e^{is(tF_1+F_2)},F_1]
e^{i(1-s)(tF_1+F_2)}\right.\right.\\
&&\left.\left. -ise^{i(tF_1+F_2)}[F_2,F_1]\right\}e^{-\frac{t}{2}[F_1,F_2]}\right).
\end{eqnarray*}
Denote
\[
A=\int_0^1 ds\left\{[e^{is(tF_1+F_2)},F_1]
e^{i(1-s)(tF_1+F_2)}-ise^{i(tF_1+F_2)}[F_2,F_1]\right\}.
\]
\begin{eqnarray*}
f'(t)&=&\omega\left(ie^{-iF_2}e^{-itF_1}A e^{-\frac{t}{2}[F_1,F_2]}\right)\\
&=&\omega\left(ie^{-iF_2}e^{-itF_1}e^{-\frac{t}{2}[F_1,F_2]}A +
ie^{-iF_2}e^{-itF_1}[A, e^{-\frac{t}{2}[F_1,F_2]}]\right),
\end{eqnarray*}
then
\beq\label{f'}
|f'(t)|\leq \left|\omega\left(e^{-iF_2}e^{-itF_1}e^{-\frac{t}{2}[F_1,F_2]}A\right)
\right|+\left|\omega\left(e^{-iF_2}e^{-itF_1}[A, e^{-\frac{t}{2}[F_1,F_2]}]\right)\right|.
\eeq
The first part is estimated as follows:
\[
\left|\omega\left(e^{-iF_2}e^{-itF_1}e^{-\frac{t}{2}[F_1,F_2]}A\right)\right|\leq 
\|A\|_\omega,
\]
by the Cauchy-Schwarz inequality. We now make an estimation of $\|A\|_\omega$:
\begin{eqnarray*}
\|A\|_\omega &=& \left\|\int_0^1 ds\left\{[e^{is(tF_1+F_2)},F_1]
e^{i(1-s)(tF_1+F_2)}-ise^{i(tF_1+F_2)}[F_2,F_1]\right\}\right\|_\omega\\
&\leq&\int_0^1 ds \left\| [e^{is(tF_1+F_2)},F_1]
e^{i(1-s)(tF_1+F_2)}-ise^{i(tF_1+F_2)}[F_2,F_1] \right\|_\omega\\
&=&\int_0^1 ds \left\|\int_0^1dr e^{irs(tF_1+F_2)}is[F_2,F_1]e^{i(1-r)s(tF_1+F_2)}
e^{i(1-s)(tF_1+F_2)} \right.\\
&&\left.-ise^{i(tF_1+F_2)}[F_2,F_1] \right\|_\omega\\
&=&\int_0^1 ds \left\|\int_0^1dr e^{irs(tF_1+F_2)}is[F_2,F_1]e^{i(1-rs)(tF_1+F_2)}
\right.\\
&&\left.-ise^{i(tF_1+F_2)}[F_2,F_1] \right\|_\omega\\
&=&\int_0^1 ds \left\|\int_0^1dr ise^{irs(tF_1+F_2)}\left\lbrack[F_2,F_1],
e^{i(1-rs)(tF_1+F_2)}\right\rbrack\right\|_\omega\\
&\leq&\int_0^1 ds\int_0^1dr s\left\|\left\lbrack[F_2,F_1],
e^{i(1-rs)(tF_1+F_2)}\right\rbrack\right\|_\omega\\
&\leq&\int_0^1 ds\int_0^1dr s(1-rs)\left\|\left\lbrack[F_2,F_1],
tF_1+F_2\right\rbrack\right\|_\omega\\
&=&\frac{1}{3}\left\|\left\lbrack[F_2,F_1],
tF_1+F_2\right\rbrack\right\|_\omega,
\end{eqnarray*}
where we have used (\ref{dys-com1}) in the third step and (\ref{dys-com2}) in the fifth.
But because the commutator $[F_2,F_1]$ converges to a complex number in the state
$\omega$, it can be calculated explicitly that the commutator
$[[F_2,F_1],tF_1+F_2]$ converges to $0$. By a similar argument it can also be shown that
the second part of (\ref{f'}) converges to $0$, thus proving Proposition \ref{BCH-prop}.
\qed

\section{Proof of Proposition \ref{cltprop}}\label{CLT-proof}
Take some arbitrary $F_{L,q}(f,g)$ and first of all remark that it can be
written as
\[
F_{L,q}(f,g) = B^*_q + B_q,
\]
where $B^*_q=B_{-q}$ and $[B^*_q,B_q]=0$ (we have only written the
$q$-dependence explicitly, all other dependencies are implicit). One can for
instance take
\[
B_{q}=\frac{1}{2(\rho_0V)^{1/2}}\sum_k f(k+q,k)a^*_{L,k+q}a_{L,k}
+\frac{i}{2}\left( g(q,0) a^*_{L,q} - g(0,q) a_{L,-q} \right).
\]
 $B_{q}$ itself can be decomposed into
\beq\label{Bq-B0}
B_q=B^0_q + \bar B_q,
\eeq
with
\begin{eqnarray*}
B_q^0 &=& \frac{1}{2(\rho_0V)^{1/2}}\left( f(q,0)a^*_{L,q}a_{L,0}
+f(0,q)a^*_{L,0}a_{L,-q} \right) \\
&&+ \frac{i}{2}\left( g(q,0) a^*_{L,q} - g(0,q) a_{L,-q} \right)\\
&=&\frac{1}{2}\left[ \left( f(q,0)\frac{a_{L,0}}{(\rho_0V)^{1/2}} + i g(q,0)
\right) a^*_{L,q} \right.\\
&& \left.+ \left( f(0,q)\frac{a^*_{L,0}}{(\rho_0V)^{1/2}} - i g(0,q)
\right)a_{L,-q} \right].
\end{eqnarray*}

We want to show now that the part $(B_q^0)^*+B_q^0$ is the only part of 
$F_{L,q}(f,g)$ which gives a non-zero contribution to the expectation value on
the l.h.s. of (\ref{clt}). Expanding the exponential in a power series, we have
to calculate expectation values of the type
\[
\omega\left( (B^*_q + B_q)^m \right).
\]
This is obviously zero for $m$ odd, and for $m=2n$ even, the only non-zero terms
are those with a number of starred operators equal to unstarred. Because of the
commutation $[B^*_q,B_q]$, these are all equal to
\[
\omega\left( (B^*_q)^nB_q^n \right).
\]
Using the decomposition (\ref{Bq-B0}) this becomes
\[
\omega\left( (B^*_q)^nB_q^n \right)=
\omega\left( ((B_q^0)^*)^n(B^0_q)^n \right) + \mathrm{other\ terms}.
\]
We now prove that these `other terms' are all necessarily zero because we are
 working in the ground state. Using the commutation of the $(B^0_q)^\sharp$ with
the $(\bar B_q)^\sharp$ these terms are all of the form
\beq\label{0*0-q*q}
\omega\left( (B_q^0)^*)^i(B^0_q)^j (\bar B_q^*)^{n-i}(\bar B_q)^{n-j} 
\right).
\eeq
We have
\begin{eqnarray*}
(\bar B_q)^{n-j} = \frac{1}{(4\rho_0V)^{\frac{n-j}{2}}}
\sum_{k_1,\ldots,k_{n-j}; -q\not=k_l\not=0} && f(k_1+q,k_1)
\ldots f(k_{n-j}+q,k_{n-j}) \\
&&a^*_{L,k_1+q}a_{L,k_1}\ldots a^*_{L,k_{n-j}+q}a_{L,k_{n-j}}.
\end{eqnarray*}
Using this in (\ref{0*0-q*q}) one gets sums of expectation values which can be
computed by using the quasi-freeness of the state $\omega$, i.e. each
expectation is a sum of products of one- and two-point correlations, respectivily
$\omega(a^\sharp_{L,k})$ and 
\[
\omega^T(a^\sharp_{L,k}a^\sharp_{L,k'})=\omega(a^\sharp_{L,k}
a^\sharp_{L,k'})-\omega(a^\sharp_{L,k})\omega(a^\sharp_{L,k'}),
\]
of which only the following are non-zero:
\begin{eqnarray*}
\omega(a^\sharp_{L,0})&=&\sqrt{\rho_0V}\\
\omega^T(a_{L,k}a^*_{L,k})&=&\omega(a_{L,k}a^*_{L,k})=1, \;\;
k\not=0.
\end{eqnarray*}
We show now that each of the expectations to be computed is zero. Take an
arbitrary term. 
First of all, take some $a_{L,0}a^*_{L,q}$ term from one of the $B^0_q$'s. 
It can only give a non-zero
contribution if it is combined with an $a^*_{L,0}a_{L,q}$ term from one of the
$(B^0_q)^*$'s, so at least we need $i=j$. A term $a^\sharp_{L,\pm q}$ from any of
the $B^0_q$'s or $(B^0_q)^*$'s can never be combined with a term coming from the
$(\bar B_q)^\sharp$'s because this would give rise to an expectation of an odd
number of creation and annihilation operators, all with a non-zero index, which is
zero. Hence in the quasi-free decomposition, the operators coming from
$((B_q^0)^*)^j(B^0_q)^j$ and the operators coming from
$(\bar B_q^*)^{n-j}(\bar B_q)^{n-j}$ completely decouple from each other. But a
typical factor arising from $(\bar B_q^*)^{n-j}(\bar B_q)^{n-j}$ is
\[
\omega\left(a^*_{L,k_1+q}a_{L,k_1}\ldots a^*_{L,k_{n-j}+q}a_{L,k_{n-j}}
a^*_{L,l_1-q}a_{L,l_1}\ldots a^*_{L,l_{n-j}+q}a_{L,l_{n-j}} \right),
\]
which is zero because it always contains at least one factor 
$\omega(a^*_{L,k}a_{L,k})$, $k\not=0$.

Hence we proved that indeed 
\[
\omega\left( (B^*_q)^nB_q^n \right)=
\omega\left( ((B_q^0)^*)^n(B^0_q)^n \right) ,
\]
or in other words
\beq\label{rho - rho0}
\lim_{L\to\infty}\omega\left( e^{itF_{L,q}(f,g)} \right)
=\lim_{L\to\infty}\omega\left( e^{itF^0_{L,q}(f,g)} \right),
\eeq
where
\[
F^0_{L,q}(f,g)=\rho_{L,q}^0(f)+A_{L,q}(g),
\]
with
\[
\rho_{L,q}^0(f) = \frac{1}{2}\left[ \frac{f(q,0)}{(\rho_0V)^{1/2}}a_{L,0}
\left(a^*_{L,q}+a^*_{L,-q}\right)+ \frac{f(0,q)}{(\rho_0V)^{1/2}} a^*_{L,0}
\left(a_{L,q}+a_{L,-q}\right)\right].
\]
Hence
\begin{eqnarray*}
F^0_{L,q}(f,g) &=& \frac{1}{2}\left\{ \left[f(q,0)\frac{a_{L,0}}{(\rho_0V)^{1/2}}
+ig(q,0) \right]\left(a^*_{L,q}+a^*_{L,-q}\right) \right.\\
&&\left. +\left[f(0,q)\frac{a^*_{L,0}}{(\rho_0V)^{1/2}}
-ig(0,q) \right] \left(a_{L,q}+a_{L,-q}\right) \right\}.
\end{eqnarray*}

Using (\ref{a0}) one gets
\[
\lim_{L\to\infty}\omega\left( e^{itF^0_{L,q}(f,g)} \right)=
\]
\beq\label{rho000} 
\lim_{L\to\infty}\omega\left( e^{\frac{it}{2}\left\{ \left[f(q,0)
+ig(q,0) \right]\left(a^*_{L,q}+a^*_{L,-q}\right) 
+ \overline{\left[f(q,0)+ig(q,0) \right]} \left(a_{L,q}+a_{L,-q}\right) \right\}}
\right).
\eeq 
With the explicit expression of the state (\ref{state}),
the r.h.s. of (\ref{rho000})  can easily be computed, and together with
(\ref{rho - rho0}) one gets (\ref{clt}).  
\qed

\section{Proof of Proposition \ref{recthm}}\label{recthm-proof}
We prove (\ref{clt-expl}) by induction on $n\in\N_0$. The basis of the induction is
the Central Limit Theorem, Proposition \ref{cltprop}.

To prove the induction step, assume that (\ref{clt-expl}) holds for some $n\in\N_0$
and fix $(f_i,g_i) \in (\mathcal{F},\mathcal{F}), 
i=1,\ldots,n+1$.

For convenience write
\[
 e^{iF_{L,q}(f_1,g_1)}\cdots e^{iF_{L,q}(f_{n-1},g_{n-1})} \equiv W_{L,q},
\] 
and
\[
F_{L,q}(f_i,g_i) \equiv F_{L,q}^i.
\]
By the Cauchy-Schwarz inequality and the BCH-formula (\ref{BCH})
\begin{eqnarray*}
&&\lim_{L\to\infty}\left| \omega\left(W_{L,q}\left[ e^{iF_{L,q}^n}
e^{iF_{L,q}^{n+1}} - e^{i(F_{L,q}^n+F_{L,q}^{n+1})}
e^{-\frac{1}{2}[F_{L,q}^n,F_{L,q}^{n+1})]} \right]  \right) \right|\\
&\leq& \lim_{L\to\infty}\left\|e^{iF_{L,q}^n}
e^{iF_{L,q}^{n+1}} - e^{i(F_{L,q}^n+F_{L,q}^{n+1})}
e^{-\frac{1}{2}[F_{L,q}^n,F_{L,q}^{n+1})]}\right\|_\omega =0.\\
\end{eqnarray*}
Use the Cauchy-Schwarz inequality again to derive that
\begin{eqnarray*}
&&\left| \omega\left(W_{L,q}e^{i(F_{L,q}^n+F_{L,q}^{n+1})}
e^{-\frac{1}{2}[F_{L,q}^n,F_{L,q}^{n+1})]} \right) -
 \omega\left(W_{L,q}e^{i(F_{L,q}^n+F_{L,q}^{n+1})}\right) 
e^{-\frac{i}{2}\sigma_q(n|n+1)} \right|^2\\
&&=\left| \omega\left(W_{L,q}e^{i(F_{L,q}^n+F_{L,q}^{n+1})}
\left[ e^{-\frac{1}{2}[F_{L,q}^n,F_{L,q}^{n+1})]}  - 
e^{-\frac{i}{2}\sigma_q(n|n+1)}\mathbf{1}\right] \right) \right|^2\\
&&\leq \omega\left(
\left[ e^{\frac{1}{2}[F_{L,q}^n,F_{L,q}^{n+1})]}  - 
e^{\frac{i}{2}\sigma_q(n|n+1)}\mathbf{1}\right] 
\left[ e^{-\frac{1}{2}[F_{L,q}^n,F_{L,q}^{n+1})]}  - 
e^{-\frac{i}{2}\sigma_q(n|n+1)}\mathbf{1}\right]\right) \\
&&=2 - \omega\left(e^{\frac{1}{2}[F_{L,q}^n,F_{L,q}^{n+1})]}\right)
e^{-\frac{i}{2}\sigma_q(n|n+1)} -
\omega\left(e^{-\frac{1}{2}[F_{L,q}^n,F_{L,q}^{n+1})]}\right)
e^{\frac{i}{2}\sigma_q(n|n+1)}.
\end{eqnarray*}
This expression converges to zero as $L\to\infty$ because of the convergence of
the commutator. Combining the induction hypothesis and the above result, one finds that
\begin{eqnarray*}
\lim_{L\to\infty}\omega\left(W_{L,q} e^{iF_{L,q}^n}
e^{iF_{L,q}^{n+1}}\right) &=& \lim_{L\to\infty}\omega\left(W_{L,q}
e^{i(F_{L,q}^n+F_{L,q}^{n+1})}\right)e^{-\frac{i}{2}\sigma_q(n|n+1)}\\
&=& \tilde \omega^q \Bigl( W_q(1) \cdots [W_q(n)+W_q(n+1)]\Bigr)\\
&&e^{-\frac{i}{2}\sigma_q(n|n+1)}\\
&=& \tilde \omega^q \left( W_q(1) \cdots W_q(n)W_q(n+1)\right).
\end{eqnarray*}
The last equality results from the $CCR$ algebraic structure of\\ $CCR((\mathcal{F},
\mathcal{F}),\sigma_q)$.

The only thing left to prove is positivity. Take $f_i,g_i \in\mathcal{F}, i=1,2$ and
use the definitions of $\sigma_q$ and $s_q$
along with the Cauchy-Schwarz inequality to derive that
\begin{eqnarray*}
\frac{1}{4}\left| \sigma_q(f_1,g_1|f_2,g_2)\right|^2 &\leq& \left| 
<f_1,g_1|f_2,g_2>_q\right|^2 \\
\leq <f_1,g_1|f_1,g_1>_q <f_2,g_2|f_2,g_2>_q &\leq& s_q(f_1,g_1|f_1,g_1)
s_q(f_2,g_2|f_2,g_2).
\end{eqnarray*}
\qed

\section{Proof of Proposition \ref{equiv-prop}}\label{equiv-proof}
Denote 
\[
W(f_i,g_i)\equiv W_i.
\]
Suppose first that (ii) is satisfied, then
\[
\left\lbrack\pi_{\tilde \omega}\left( W_1 \right),
\pi_{\tilde \omega}\left( W_2 \right)\right\rbrack=0
\]
and hence
\[
\sigma(1|2)=0.
\]
Further
\begin{eqnarray*}
1 &=& \tilde\omega\left( W_1W_2^*\right) = \tilde
\omega\left( W_1W_{-2}\right)\\
&=& \tilde\omega\left( W_{1-2} \right) = e^{-\frac{1}{2}s(1-2|1-2)},
\end{eqnarray*}
where we used the notation $(f_1-f_2,g_1-g_2) \to 1-2$.
By the definition of $s$ this means $<1-2|1-2>=0$, proving (i).

Conversely, suppose $(f_1,g_1) \sim (f_2,g_2)$ then
\[
\frac{1}{4}\left| \sigma(1-2|x)\right|^2 \leq <1-2|1-2> <x|x>
\]
implies that $\sigma(1-2|x)=0$ for all $x$, where $x$ denotes an arbitrary element of
$(\mathcal{F},\mathcal{F})$, i.e. $\pi_{\tilde \omega}\left( W_{1-2} \right)$
commutes with all elements of $\pi_{\tilde \omega}(CCR((\mathcal{F},\mathcal{F}),\sigma))
\equiv\mathcal{M}$, or $\pi_{\tilde \omega}\left( W_{1-2} \right)$ belongs to the
commutant $\mathcal{M}'$ of $\mathcal{M}$.
Also
\begin{eqnarray*}
\left\| \left( \pi_{\tilde \omega}\left( W_{1-2} \right)-
\mathbf{1} \right) \Omega_{\tilde \omega} \right\|^2 &=&
\tilde \omega\left((W_{1-2}-\mathbf{1})^*(W_{1-2}-\mathbf{1}) \right)\\
&=& 2 - \omega\left((W_{1-2}\right) - \omega\left((W_{2-1}\right)\\
&=& 0.
\end{eqnarray*}
As $\Omega_{\tilde \omega}$ is cyclic for $\mathcal{M}$ it is separating for 
$\mathcal{M}'$. Hence
\[
\pi_{\tilde \omega}\left( W_{1-2} \right) = \mathbf{1}
\]
or
\[
\pi_{\tilde \omega}\left( W_{1} \right)=\pi_{\tilde \omega}
\left( W_{2} \right).
\]
\qed

\end{document}